\documentclass[reprint,superscriptaddress,aps,pra]{revtex4-2}
\usepackage[T1]{fontenc}
\usepackage[latin1]{inputenc}
\usepackage{lmodern}
\expandafter\let\csname equation*\endcsname\relax
\expandafter\let\csname endequation*\endcsname\relax
\usepackage{amsmath}
\usepackage{amssymb,dsfont}
\usepackage{graphicx}
\usepackage{xcolor}
\usepackage{natbib}
\usepackage{bm}
\usepackage{bbold}
\usepackage[mathscr]{euscript}
\usepackage[cal=boondoxo]{mathalfa}
\usepackage{comment}
\renewcommand\footnotemark{}

\usepackage{braket}
\usepackage[normalem]{ulem}
\usepackage{lineno}

\def\ii{{\rm i}} 

\begin{document}

\title{Fast entanglement generation via cavity-mediated nematic tensor dynamics}
\author{Stuart J. Masson}
\email{sjmasson@usf.edu}
\affiliation{Department of Physics, University of South Florida, Tampa, Florida 33620, USA}

\date{\today}

\begin{abstract}
Many-body entanglement in atomic ensembles offers the potential for enhanced sensing devices. Entanglement can be generated by interfacing atoms with a cavity. Traditional methods use effective two-level systems isolated from more complex atomic level structures. However, alternative strategies allow entanglement to be generated in atoms confined to a single hyperfine manifold. These strategies enable entanglement in ways fundamentally not available to two-level systems but feature extremely slow dynamics due to requiring strongly off-resonant laser fields. Here, I show that equivalent squeezing dynamics can be achieved much closer to resonance, offering entanglement generation over a hundred times faster than typical experiments. Furthermore, the Hamiltonian can be used to speed up the creation of maximally entangled states via adiabatic passage.
\end{abstract}

\maketitle

\section{Introduction}

Interfacing large atomic ensembles with optical cavities provides a flexible platform for the study of many-body systems strongly coupled to light and to engineer interactions between atoms~\cite{Ritsch13,Mivehvar21}. A prominent application is to generate spin squeezing interactions between atoms~\cite{Kitagawa93,Ma11,Pezze18}. By reducing quantum noise in one spin component at the expense of another, spin squeezed states offer metrological sensitivity with variance bypassing the standard quantum limit set by uncorrelated atoms of $1/N$~\cite{Wineland94}. Spin squeezed states are most commonly generated in cavities via quantum non-demolition measurement and cavity feedback~\cite{Leroux10PRL1,Cox16,Hosten16}. Recent demonstrations of spin squeezing on clock transitions~\cite{Pedrozo20,Robinson24} show that these technologies are close to offering real-world advantage.

Spin-squeezing interactions can be engineered via multi-photon transitions~\cite{Sorensen02}. This approach limits decoherence because atoms are limited to stable ground states, while adding flexibility to interactions between the atoms and the cavity as they become dressed by the presence of the external drives. Such schemes have helped place many-body cavity QED systems at the forefront of quantum science, including demonstrations of self-organizing phase transitions~\cite{Black03,Baumann10,Klinder15,Leonard17Nature,Zhiqiang18,Kroeze18,Kongkhambut22,Ho25}, interaction engineering~\cite{Dogra16,Vaidya18,Periwal21,Marsh24,Young24}, and creation of novel light sources~\cite{Bonifacio98,Slama07,Bohnet12,Groiseau21}. The integration of tweezer arrays into cavity QED has generated further interest in this area of research as it provides a microscopically controllable many-body light-matter interaction platform~\cite{Deist22,Liu23,Zhenjie23,Hartung24,Ho25,DeSantis26arxiv}.

Many-body cavity QED experiments usually isolate a two-level system for each atom formed either of two motional states~\cite{Nagy10} or two electronic states~\cite{Dimer07}. This creates an ensemble of atoms each described by a spin-1/2 algebra. If $N$ atoms are coupled identically to the cavity, permutational symmetry allows summation of the individual angular momenta and restricts the atoms to a single angular momentum ladder of length $N/2$. The majority of many-body cavity QED experiments are operated in this limit. However, restriction of each atom to two levels is either an approximation that only holds for short times or must be enforced by specific choices of light polarization or repumping to offset leakage.

A natural level structure to consider in alkali atoms are entire ground-state hyperfine manifolds. This provides a set of near-degenerate states without any other states with nearby frequencies, alleviating leakage concerns. Naturally, this approach goes beyond the two-level system and, in general, requires multilevel systems with a total angular momentum $F$ to describe the system. The interaction of identical spin-$F$ atoms with a cavity mode is still described solely by spin angular momentum operators in the limit that the detuning of the cavity and laser fields is much larger than the hyperfine splitting of the excited states mediating the multiphoton transitions~\cite{Masson17,Zhiqiang17,Davis19}. If the atoms are initiated in fully-polarized states, i.e. $\ket{F,m_F=\pm F}$, summation over the atoms still forms a single angular momentum ladder, now of length $NF$, and allows for equivalent physics to the spin-1/2 case~\cite{Zhiqiang17}. However, initiating atoms outside this state leads to initial states with a superposition of total spin~\cite{Masson19PRL} and enables entanglement to be generated between atoms in a manner fundamentally not available to spin-1/2 atoms: spin-nematic squeezing~\cite{Duan02,Hamley12,Masson17,Cooper24}. This is an analogous process to spin squeezing in spin-1/2 particles, but where noise redistribution occurs within an operator space exclusive to larger spin particles.

The requirement of such a large detuning is inconvenient. The interaction rate scales as the inverse of the detuning and so large laser power is required or dynamics are extremely slow. Accessing the novel physics available to spinor particles motivates the need for alternative approaches that do not require such an extreme detuning. In this manuscript, I show that entanglement can be generated in ensembles of $^{87}$Rb atoms driven much closer to resonance. In contrast to the large detuning limit, where angular momentum operators describe interactions, I present a detuning condition where the angular momentum operator terms are precisely zero and interactions are exclusively mediated by operators specific to describing spin-1 atoms. I show that these terms can be used to generate spin-nematic squeezing and facilitate adiabatic preparation of entangled ground states in spinor Bose-Einstein condensates. Since this detuning condition is much closer to resonance, it enables entanglement generation in spin-1 atoms on timescales over a hundred times faster than current methods.

\section{Atom-cavity model}

\begin{figure}[t!]
    \includegraphics[width=\columnwidth]{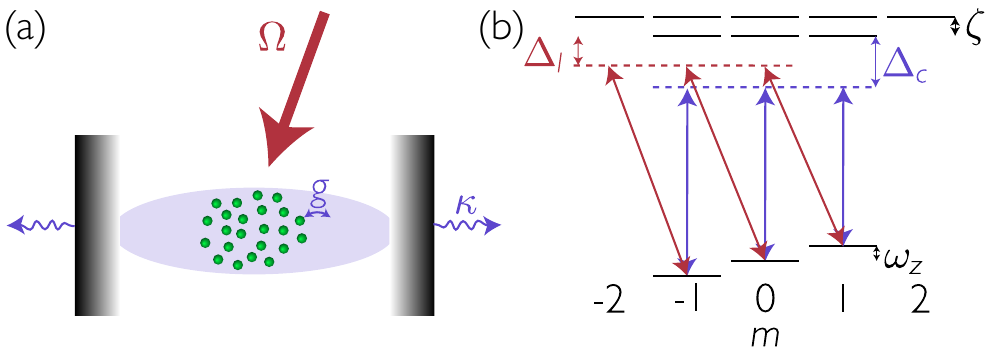}
    \caption{Setup. (a) $^{87}$Rb atoms are driven with Rabi frequency $\Omega$ and coupled at strength $g$ to an optical cavity which decays at $\kappa$. (b) The laser (cavity) fields are off-resonant from the $D_1$ line by $\Delta_{l(c)} = \omega_{l(c)} - \omega_1 = \omega_{l(c)} - \omega_2 + \zeta$ where $\omega_{1(2)}$ are the frequencies of the $F=1(2)$ line and $\zeta$ is their splitting.\label{fig:setup}}
\end{figure}

I consider an ensemble of $^{87}$Rb atoms in an optical cavity off-resonant from the $D_1$ line, as shown in Fig.~\ref{fig:setup}. Atoms are confined to the $F=1$ ground state manifold with transitions between their states driven by a classical $\sigma_-$-polarized laser field and a $\pi$-polarized cavity mode. The system is described by the master equation ($\hbar=1$)
\begin{equation}
\dot{\rho} = - \ii \left[\sum\limits_{j=1}^N \hat{H}_{\mathrm{atom}}^j + \hat{H}_\mathrm{int.}^j,\rho\right] + \kappa \mathcal{D}[\hat{a}]\rho.
\end{equation}
Here, $\hat{a}$ and $\kappa$ are the annihilation operator and decay rate of the cavity mode and the sum runs over all atoms. Loss is included via $\mathcal{D}[\hat{a}]\rho = 2 \hat{a} \rho \hat{a}^\dagger - \{ \hat{a}^\dagger\hat{a} ,\rho\}$ and atomic decay is neglected. The single atom Hamiltonian is
\begin{equation}
    \hat{H}_{\mathrm{atom}} = \sum\limits_{m = -1}^{1} m\omega_{z} \ket{m} \bra{m} + \sum\limits_{F=1,2}\sum\limits_{f = -F}^{F} \omega_{F} \ket{F,f}\bra{F,f},
\end{equation}
where $\ket{m}$ denotes the ground state with magnetic number $m$ and $\ket{F,f}$ denotes excited states in the $F$ hyperfine level with magnetic number $f$, $\omega_z$ is the linear Zeeman shift of the ground state manifold, and $\omega_F$ is approximated as the frequency of the $F$ excited manifold for all $f$, neglecting Zeeman shifts. The single atom interaction Hamiltonian is given in the interaction picture by
\begin{align}
 \hat{H}_{\mathrm{int.}} = &\sum\limits_{m,F,f} g \mathrm{e}^{-\ii\omega_c t} \hat{a} d_{m,0}^F \ket{F,f}\bra{m}\notag \\&+ \frac{\Omega}{2} \mathrm{e}^{-\ii\omega_l t} d_{m,-1}^F \ket{F,f-1}\bra{m} + \mathrm{h.c.},
 \end{align}
 where $g$ is the coupling strength to the cavity mode, $\Omega$ is the Rabi frequency of the laser, $\omega_{c,l}$ are the frequencies of the cavity and laser respectively, and $d^{F}_{m,q}$ is the dipole matrix element for the transition $\ket{m} \leftrightarrow \ket{F,f+q}$.

\subsection{Adiabatic elimination of the excited states}

 In the limit that $\omega_{c,l} - \omega_F \gg \Omega, g$ the  excited states may be adiabatically eliminated using a projector operator formalism~\cite{Reiter12}. The Hamiltonian is reduced to an effective form (see Appendix~\ref{app:excitedstates} for the full derivation)
    \begin{align}\label{eq:fullmodelfield}
        &\hat{H}_{\mathrm{eff.}} =  \omega_0 \hat{S}_z + \omega_q \hat{Q}_{zz} + \frac{\delta_q}{2N} \hat{Q}_{zz} \hat{a}^\dagger\hat{a} \\&+ \frac{\lambda}{\sqrt{2N}} \left( \hat{a}\hat{S}_+ + \hat{a}^\dagger\hat{S}_- \right) + \frac{\phi}{2\sqrt{2N}} \left( \hat{a}\hat{Q}_{+z} + \hat{a}^\dagger\hat{Q}_{-z} \right) \notag,
    \end{align}
where $\hat{S}_j$ are collective spin operators formed by summation over the spin operators of each atom and $\hat{Q}_{jk}$ are elements of the collective nematic tensor operator formed in the same manner with single atom operators~\cite{Hamley12}
\begin{subequations}\label{eq:nematictensordefinitions}
    \begin{align}
        \hat{q}_{zz} &= \frac{2}{3}\ket{1}\bra{1} + \frac{2}{3}\ket{-1}\bra{-1} - \frac{4}{3} \ket{0}\bra{0} \\
        \hat{q}_{+z} &= 2\sqrt{2} \left( - \ket{0}\bra{-1} + \ket{1}\bra{0} \right) = \hat{q}_{-z}^\dagger.
    \end{align}
\end{subequations}
$\hat{Q}_{zz}$ is effectively a quadratic Zeeman shift while $\hat{Q}_{\pm z}$ act as ladder operators with a relative sign difference between the two transitions. The effective parameters in Eq.~\eqref{eq:fullmodelfield} depend on the microscopic parameters of the system as
\begin{subequations}
    \begin{align}
        \omega_0 &= \omega_z + \frac{\Omega^2}{96} \left(\frac{1}{\Delta} - \frac{5}{\Delta - \zeta}\right) \\
        \omega &=  \omega_c - \omega_l + \frac{N g^2}{18}\left(\frac{1}{\Delta} + \frac{5}{\Delta-
        \zeta} \right) \label{eq:effomega} \\
        \omega_q &=  \frac{\Omega^2}{192}\left(\frac{1}{\Delta-\zeta} - \frac{1}{\Delta} \right) \\
        \delta_q &= \frac{Ng^2}{12} \left( \frac{1}{\Delta} - \frac{1}{\Delta - \zeta} \right) \\
        \lambda &= \frac{\sqrt{N}\Omega g}{48}\left(\frac{1}{\Delta} - \frac{5}{\Delta - \zeta} \right)  \\
        \phi &= \frac{\sqrt{N} \Omega g}{48} \left(\frac{1}{\Delta - \zeta} - \frac{1}{\Delta} \right).
    \end{align}
\end{subequations}
Here, I define $\Delta \equiv \omega_{c,l} - \omega_1$ as the detuning from the $F=1$ excited hyperfine manifold assuming $\omega_c\approx\omega_l$ and $\zeta \equiv \omega_2 - \omega_1$ as the excited-state hyperfine splitting.

There are two specific regimes that we will work in throughout this manuscript:
\begin{itemize}
\item \textbf{Pure spin models, $|\Delta| \gg \zeta$\label{sec:purespin}}. In the limit that the detuning from the $D_1$ line becomes much larger than the hyperfine splitting $\zeta$, the detuning is approximately the same from both hyperfine excited states and all parameters related to the nematic tensor operators go to zero~\cite{Masson17}.

\item \textbf{Pure nematic models, $\Delta = -\zeta/4$\label{sec:purenematic}}. A model with $\lambda=0$ can be achieved if $\Delta = -\zeta / 4$ and $\Delta - \zeta = -5\zeta/4$. ac Stark shift contributions to the effective atomic frequency are also zero.
\end{itemize}
\color{black}

The general form of the effective Hamiltonian in Eq.~\eqref{eq:fullmodelfield} can be generated with other atomic species with different numerical factors in the effective parameters due to different Clebsch-Gordan coefficients. I restrict myself here to an $F=1$ ground state for convenience, but the same form of effective Hamiltonian is true for larger $F$. While moving from two ground states to three introduces complexity that requires the nematic tensor operators defined in Eq.~\eqref{eq:nematictensordefinitions}, going beyond three states does not introduce further complexity. This fact is due to these interactions being mediated by two photons such that terms beyond tensor rank are not possible; this is a generalization of the polarizability only having up to tensor rank terms. The same entangling dynamics described in the following hold for $^{87}$Rb atoms initialized in the $F=2$ manifold or for other alkali atoms.

\subsection{Adiabatic elimination of the cavity mode}

Atom-atom interactions are obtained in parameter regimes where the cavity mode can also be adiabatically eliminated. If the effective cavity frequency $\omega$ in Eq.~\eqref{eq:effomega} or cavity decay rate $\kappa$ are much larger than the effective cavity couplings $\lambda,~\phi$, an effective master equation governing only the atoms can be derived~\cite{Sorensen02,Morrison08PRL,Masson17}. Since photon population is negligible, terms proportional to $\hat{a}^\dagger\hat{a}$ are ignored and the Hamiltonian is approximated well by
\begin{align}
    \hat{H} \approx \omega_0 \hat{S}_z + \omega_q \hat{Q}_{zz} + \hat{a}\hat{X}_+ + \hat{a}^\dagger\hat{X}_-,
\end{align}
where I have defined the operators 
\begin{equation}
 \hat{X}_\pm = \frac{\lambda}{\sqrt{2}} \hat{S}_\pm + \frac{\phi}{2\sqrt{2}} \hat{Q}_{\pm z}.
\end{equation}
Adiabatic elimination of the cavity mode generates the effective master equation for the atomic state~\cite{Masson17}
\begin{subequations}\label{eq:genericHamiltonianbeforedetuning}
\begin{align}
    \dot{\rho} &= - \ii \left[\hat{H},\rho\right] + \frac{\kappa}{N(\kappa^2 + \omega^2)}\mathcal{D}[\hat{X}_+]\rho \\
    \hat{H}_{\mathrm{eff.}} &= \omega_0\hat{S}_z + \omega_q \hat{Q}_{zz} - \frac{\omega}{N(\kappa^2 + \omega^2)}\hat{X}_-\hat{X}_+. 
\end{align}
\end{subequations}

\section{Bosonic operator representation}

To understand why the spin models induce entanglement, I make use of a second quantization of the atomic states. In this form, the relevant atomic operators are
\begin{subequations}
\begin{align}
    \hat{S}_z &= \hat{b}^\dagger_1\hat{b}_1 - \hat{b}^\dagger_{-1}\hat{b}_{-1} \\
    \hat{Q}_{zz} &= \frac{2}{3} \left( \hat{b}_1^\dagger\hat{b}_1 + \hat{b}_{-1}^\dagger\hat{b}_{-1} \right) - \frac{4}{3} \hat{b}_0^\dagger\hat{b}_0 \\
    \hat{S}_+ &= \sqrt{2}\left(\hat{b}^\dagger_{1}\hat{b}_0 + \hat{b}^\dagger_0\hat{b}_{-1}\right) \\
    \hat{Q}_{+z} &= 2\sqrt{2} \left( \hat{b}^\dagger_1\hat{b}_0 - \hat{b}_0^\dagger\hat{b}_{-1} \right),
    \end{align}
\end{subequations}
where $\hat{b}_j$ is an annihilation operator for an atom in state $j$. In this form, the relevant contributions to Eq.~\eqref{eq:genericHamiltonianbeforedetuning} are
\begin{subequations}
\begin{align}
    \hat{X}_+ &= (\lambda + \phi)\hat{b}^\dagger_{1}\hat{b}_0 + (\lambda - \phi) \hat{b}^\dagger_0\hat{b}_{-1} \\
    \hat{X}_-&\hat{X}_+ = \notag \left(\lambda^2 - \phi^2\right) \left(\hat{b}_1^\dagger \hat{b}_{-1}^\dagger\hat{b}_0\hat{b}_0 + \hat{b}_1\hat{b}_{-1}\hat{b}_0^\dagger\hat{b}_0^\dagger\right) \\&+ \left( \lambda + \phi\right)^2 \hat{b}_0^\dagger\hat{b}_0 \hat{b}_1\hat{b}_1^\dagger + (\lambda - \phi)^2 \hat{b}_0\hat{b}_0^\dagger \hat{b}_{-1}^\dagger\hat{b}_{-1}.\label{eq:xmxp}
\end{align}
\end{subequations}
The first term in Eq.~\eqref{eq:xmxp} is responsible for the correlated pair creation that leads to squeezing~\cite{Hamley12}, while the latter two terms are frequency shifts on different states.

\subsection{Spin model}

In the very large detuning limit $\phi, \omega_q \approx 0$ such that
\begin{widetext}
\begin{equation}
    \hat{H} \approx \left(\omega_0 + \frac{\Lambda}{2N}\right) \left(\hat{b}_1^\dagger\hat{b}_1 - \hat{b}_{-1}^\dagger\hat{b}_{-1}\right) + \frac{\Lambda}{N} \left[\hat{b}_1^\dagger \hat{b}_{-1}^\dagger\hat{b}_0\hat{b}_0 + \hat{b}_1\hat{b}_{-1}\hat{b}_0^\dagger\hat{b}_0^\dagger + \hat{b}_0^\dagger\hat{b}_0 \left( \frac{1}{2} + \hat{b}_1^\dagger\hat{b}_1 + \hat{b}_{-1}^\dagger \hat{b}_{-1} \right)\right]\label{eq:spinsqueezing}
\end{equation}
\end{widetext}
where a constant energy shift is omitted and
\begin{equation}
    \Lambda = -\frac{\lambda^2\omega}{\kappa^2 + \omega^2}.
\end{equation}
The cavity elimination produces an effective linear Zeeman shift to the atoms, which can be compensated by appropriate choice of $\omega_0$. With that choice, the Hamiltonian is proportional to $\hat{S}_x^2 + \hat{S}_y^2$ and is the spin-nematic squeezing Hamiltonian~\cite{Duan02,Hamley12,Masson17}.

\subsection{Nematic model\label{sec:nematicmixing}}

I now consider the condition $\Delta = - \zeta / 4$. At this detuning $\lambda=0$ and the Hamiltonian becomes
\begin{widetext}
\begin{equation}
    \hat{H} = \left(\omega_0 - \frac{\Phi}{2N}\right) \left(\hat{b}_1^\dagger\hat{b}_1 - \hat{b}_{-1}^\dagger\hat{b}_{-1}\right) - 2\omega_q\hat{b}_0^\dagger\hat{b}_0 + \frac{\Phi}{N} \left[\hat{b}_1^\dagger \hat{b}_{-1}^\dagger\hat{b}_0\hat{b}_0 + \hat{b}_1\hat{b}_{-1}\hat{b}_0^\dagger\hat{b}_0^\dagger - \hat{b}_0^\dagger\hat{b}_0 \left( \frac{1}{2} + \hat{b}_1^\dagger\hat{b}_1 + \hat{b}_{-1}^\dagger \hat{b}_{-1} \right)\right]. \label{eq:nematicsqueezing}
\end{equation}
\end{widetext}
where I again discard identity terms and
\begin{equation}
    \Phi = \frac{\phi^2\omega}{\kappa^2 + \omega^2}.
\end{equation}
This nematic Hamiltonian is very similar to the spin Hamiltonian Eq.~\eqref{eq:spinsqueezing}, with an effective linear Zeeman shift that can be compensated to zero and a spin mixing term that looks like $\hat{S}_x^2 + \hat{S}_y^2$ but with a relative sign difference between the pair creation and frequency shift terms. It transpires that this combination of operators is directly proportional to $\hat{Q}_{xz}^2 + \hat{Q}_{yz}^2$ and its similar form hints that this will also lead to spin-nematic squeezing. There is also an additional effective quadratic Zeeman shift that is not trivially zero. Practically, an additional detuned $\pi-$polarized laser could be used to compensate for this effect allowing it to also become a freely controlled parameter.

\section{Spin-nematic squeezing}

\begin{figure}[b!]
\begin{center}
    \includegraphics[width=0.45\textwidth]{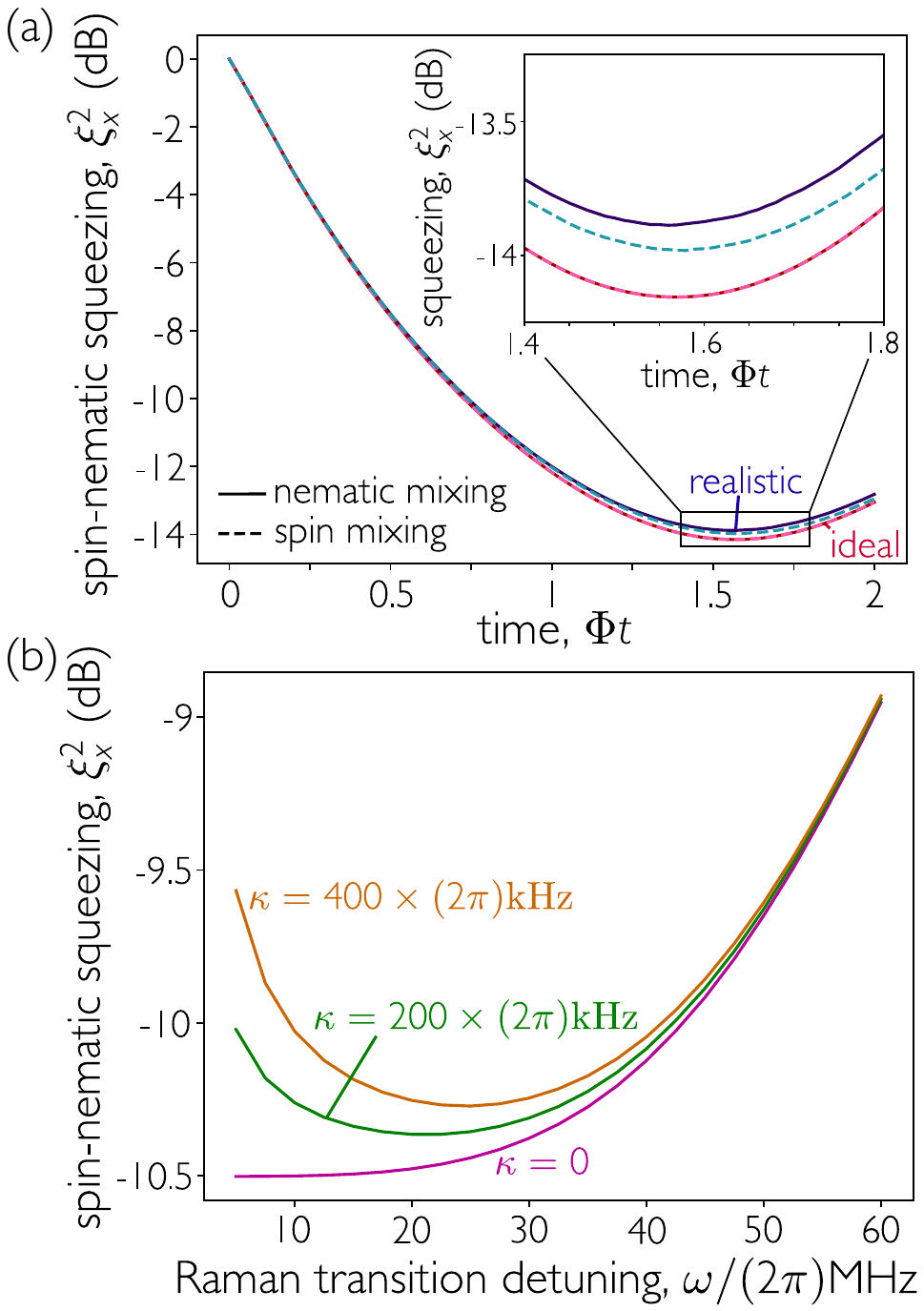}
    \caption{Spin-nematic squeezing generation. (a) Time evolution of squeezing for $N=200$ under a spin mixing Hamiltonian (dashed lines) and a nematic mixing Hamiltonian (solid lines). The ideal case evolves the initial state $\ket{\psi(t=0)} = \ket{0}^{\otimes N}$ under Eqs.~\eqref{eq:twohamss} and ~\eqref{eq:twohamsn} respectively, while the realistic case includes cavity loss at a rate $\kappa = 200 \times (2\pi)$ kHz for both cases and uses a Raman transition detuning $\omega = 20 \times (2\pi)$ MHz for nematic mixing. Lines including loss are obtained from an ensemble average of 5000 trajectories. (b) Spin-nematic squeezing via nematic mixing from an initial state $\ket{0}^{\otimes N}$ with $N=60$ as a function of Raman transition detuning for different loss values. Lines are obtained from direct evolution of the master equation.
    \label{fig:squeezing}}
    \end{center}
\end{figure}

Squeezing due to the pair creation process occurs in the SU(2) spaces defined by the spheres $\{\hat{S}_x,\hat{Q}_{yz},\hat{Q}_{zz} - \hat{Q}_{yy}\}$ and $\{\hat{S}_y,\hat{Q}_{xz},\hat{Q}_{zz} - \hat{Q}_{xx}\}$. Squeezing on each sphere is equivalent, so we consider only the first. Squeezing is quantified by the squeezing parameter~\cite{Duan02,Hamley12}
\begin{equation}
\xi_x^2 = \frac{2\mathrm{Var.}\left(\cos \theta \hat{S}_x + \sin\theta \hat{Q}_{yz}\right)}{|\braket{\hat{Q}_{zz} - \hat{Q}_{yy}}|},
\end{equation}
where $\mathrm{Var.} (\hat{\mathcal{O}}) = \braket{\hat{\mathcal{O}}^2} - \braket{\hat{\mathcal{O}}}^2$ is the variance, and the parameter $\theta$ is optimized to minimize $\xi_x^2$. 

I compare the squeezing generated by the Hamiltonians
\begin{subequations}
\begin{align}
    \hat{H}_{\mathrm{spin}} &= \frac{\Lambda}{N} \left( \hat{S}_x^2 + \hat{S}_y^2 \right) \label{eq:twohamss}\\
    \hat{H}_{\mathrm{nem.}} &= \frac{\Phi}{N} \left( \hat{Q}_{xz}^2 + \hat{Q}_{yz}^2 \right) \label{eq:twohamsn},
\end{align}
\end{subequations}
for an ensemble of spin-1 atoms initialized in the $m=0$ state. $\hat{H}_\mathrm{nem.}$ produces identical spin-nematic squeezing to $\hat{H}_\mathrm{spin}$, as shown in Fig.~\ref{fig:squeezing}(a). The only difference is that the value of $\theta$ necessary to optimize the squeezing parameter differs by $\pi/2$ for the two cases.

\subsection{Consideration of realistic parameters\label{sec:realistic}}

\begin{figure*}
\includegraphics[width=\textwidth]{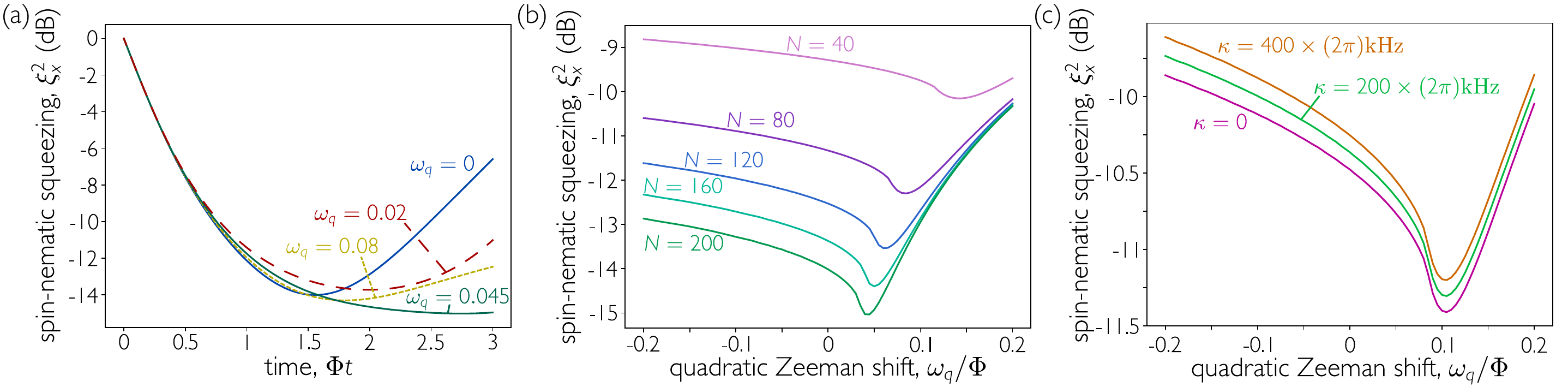}
\caption{Impact of quadratic Zeeman shift on spin-nematic squeezing. (a) Time evolution of spin-nematic squeezing for $N=200$ via nematic mixing with different quadratic Zeeman shifts. (b) Optimal spin-nematic squeezing as a function of the quadratic Zeeman shift for different $N$. (c) Optimal spin-nematic squeezing as a function of the quadratic Zeeman shift for different cavity loss levels for $N=60$. In all plots, $\omega=20\times(2\pi)$ MHz and the initial state is $\ket{0}^{\otimes N}$. In (a,b), values are calculated by evolution of the Schr\"odinger equation, while in (c) values are calculated from direct evolution of the master equation.}
\label{fig:quadzee}
\end{figure*}

The derivation of the nematic mixing Hamiltonian requires a sequence of assumptions that are not strictly compatible. In our derivation of the generic Hamiltonian, a single detuning, $\Delta$, was assumed for all fields. However, to eliminate the cavity mode we assume $\omega \gg \lambda$ which is most feasibly met by a large mismatch between the cavity and laser frequencies. Furthermore, the limit $\omega \gg \kappa$ is desirable or the evolution will be strongly dissipative and not generate squeezing. For very large detuning we can assume a hierarchy $\Delta \gg \omega_c - \omega_l \gg \kappa,\lambda$ validating the single detuning assumption. However, operation in the nematic Hamiltonian limit requires a fixed $\Delta$ such that the same hierarchy of parameters would require very low cavity loss. Relaxing the single-detuning assumption is necessary for realistic cavity parameters and requires a redefinition of the effective parameters~\cite{Reiter12}
\begin{subequations}
\begin{align*}
\lambda = \frac{g\Omega}{48} \left( \frac{\Delta_c + \Delta_l}{2\Delta_c\Delta_l} - \frac{5(\Delta_c + \Delta_l - 2\zeta)}{2(\Delta_c - \zeta)(\Delta_l - \zeta)}\right) \\
\phi = \frac{g\Omega}{96} \left(\frac{\Delta_c + \Delta_l - 2\zeta}{2(\Delta_c - \zeta)(\Delta_l - \zeta)} - \frac{\Delta_c + \Delta_l}{2\Delta_c\Delta_l}\right),
\end{align*}
\end{subequations}
where $\Delta_{c(l)}$ is the detuning of the cavity (laser) from the $D_1$ line. With the restriction $\Delta_c \neq \Delta_l$, it is not possible to set $\lambda = 0$ with $\phi \neq 0$. A balance must therefore be struck between an imperfect Hamiltonian and cavity loss. Increasing $\omega$ makes the non-pair conserving terms in the Hamiltonian larger but decreases loss. This balance leads to a broad parameter regime where substantial spin-nematic squeezing is still possible, as shown in Fig.~\ref{fig:squeezing}(b). Furthermore, large $g$ could in principle increase $\omega$ without increasing $\lambda$.

These calculations assume that spontaneous emission from the atoms to non-cavity modes is negligible and can be ignored. The effective spontaneous emission rate from an excited state is proportional to the square of the dipole matrix element and approximately proportional to the squared ratio of the Rabi frequency and the detuning from that state. Summing over all states and atoms provides an approximate spontaneous emission rate for each case
\begin{equation}
 \Gamma_{\mathrm{nem.}} \approx \frac{4\gamma\Omega^2}{5\zeta^2}, ~~~~ \Gamma_{\mathrm{spin}} \approx \frac{\gamma\Omega^2}{4\Delta^2},
\end{equation}
where $\gamma$ is the bare spontaneous emission rate of the atoms. The lower detuning associated with nematic mixing leads to a greatly enhanced spontaneous emission rate. Since the number of spontaneous emission events also depends on the amount of time required to generate the squeezing, this leads to a factor of approximately five times more spontaneous emission events in the nematic mixing case than for spin mixing.

The quadratic Zeeman shift associated with the laser drive is also not negligible in the regime $\Delta = -\zeta / 4$. To obtain the form of the nematic Hamiltonian in Eq.~\eqref{eq:twohamsn}, the quadratic Zeeman shift is assumed to be zero, implying either that $g \gg \Omega$ or that an additional quadratic Zeeman shift is applied to cancel it out. However, as shown in Fig.~\ref{fig:quadzee}, a small quadratic Zeeman shift can actually enhance the squeezing. This enhancement can be considered as a nematic version of ``twist-and-turn'' squeezing~\cite{Law01,Muessel15}, where the quadratic Zeeman shift approximates the $\hat{Q}_{zz} - \hat{Q}_{yy}$ axis of the SU(2) space and so plays the role of the transverse field and provides the additional turn. However, the induced squeezing occurs at a \emph{slower} rate than without the quadratic shift [Fig.~\ref{fig:quadzee}(a)], in contrast to twist-and-turn squeezing. This slowdown means that improvement is reduced when loss is included [Fig.~\ref{fig:quadzee}(c)], as the process taking longer provides more time for loss events to reduce the squeezing. The optimal value of $\omega_q$ is thus dependent on the emitter number and parameter regime.

\begin{figure*}[t!]
    \includegraphics[width=\textwidth]{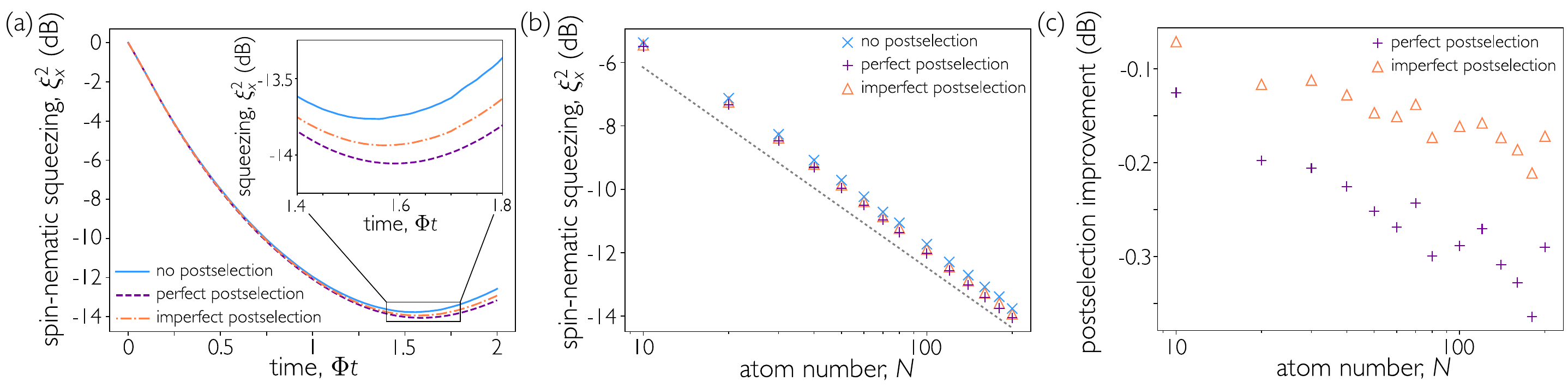}
    \caption{Scaling of spin-nematic squeezing with postselection. (a) Time evolution of spin-nematic squeezing for $N=200$. (b)~Scaling of optimal spin-nematic squeezing with $N$. The dotted black line is a guide that scales as $N^{-2/3}$. (c) Improvement of spin-nematic squeezing via postselection, i.e., $ \xi_x^2|_\mathrm{postselection} / \xi_x^2|_\mathrm{no~postselection}$. In all cases, the dataset is composed of 5000 stochastic trajectories (10000 for $N\leq120$) from an initial state $\ket{0}^{\otimes N}$. Lines and points without postselection (blue) use the ensemble average of all trajectories, those with perfect postselection (purple) use a single trajectory conditioned on no jumps, and those with imperfect postselection (orange) use an ensemble average which includes each trajectory with a probability based on photon detection efficiency of 20\%.}
    \label{fig:postselection}
\end{figure*}

\subsection{Evaluating the potential speed-up}

The power of squeezing by nematic mixing is that, whilst we strictly generate slightly less useful entanglement, we do so faster because the required detuning is substantially smaller. Assuming all other parameters are kept fixed, the ratio between the two rates is given by
\begin{equation}
    \left|\frac{\Phi}{\Lambda}\right| = \frac{16 \Delta^2}{25 \zeta^2}
\end{equation}
For the $D_1$ line of $^{87}$Rb the excited state hyperfine splitting is $\zeta = 816 \times 2\pi$~MHz~\cite{SteckRb87} and so, using $\Delta = 10 \times 2\pi$~GHz~\cite{Periwal21} for the spin mixing case, nematic mixing would provide a speedup of around 116 times. Enhancing the repetition rate by such a factor would allow substantially higher precision despite the small drop in precision per measurement. Furthermore, the speedup also reduces the time over which other experimental imperfections can degrade the squeezing, likely leading to improved single-shot precision as well.

\subsection{Postselecting optimal single-use states\label{sec:postselection}}

Loss limits the metrological sensitivity attainable. In this setup, loss is mediated by light leaking from the cavity which can in principle be collected and postselected on. Monitoring the cavity output allows for spin-nematic squeezed states conditioned on the absence of cavity emission, throwing out suboptimal contributions to the state. This leads to a substantial improvement in the sensitivity, as shown in Fig.~\ref{fig:postselection}. The total squeezing seems well-fitted to the anticipated scaling of $N^{-2/3}$~\cite{Hamley12}, as shown in Fig.~\ref{fig:postselection}(b). 

The improvement offered by postselection slowly increases with $N$, as shown in Fig.~\ref{fig:postselection}(c), albeit extracting an exact scaling is imprecise due to the stochastic nature of the generated data. Spin-nematic squeezing achieved by postselection is stronger than in the absence of loss, as the null-measurement backaction induced by the non-Hermitian Hamiltonian governing evolution enhances squeezing. It should be noted that this means that the impact of loss can be mitigated but does not mean that such loss is preferable; while postselection optimizes single-shot sensitivity, it does not improve the sensitivity of repeated measurements once the loss in efficiency is taken into consideration~\cite{Combes14}.

Inefficient photon collection still leads to improved squeezing. Only collecting a photon with 20\% efficiency still significantly improves squeezing, as shown by the orange lines and triangles in Fig.~\ref{fig:postselection}. The more photons produced in an individual instance, the more detrimental that state is to the ensemble average sensitivity, and so conditioning even on a low collection efficiency still allows for removal of the worst unravellings in the density matrix. This is compounded by runaway processes that can occur in certain instances. The initial state is a superposition of collective states with different emission rates~\cite{Masson19PRL}. Since high emission rates states are more likely to produce photons, the state following a photon emission has higher relative weight in the resultant superposition. This process feeds into itself in a manner reminiscent of Dicke superradiance~\cite{Gross82}.

\section{Adiabatic preparation of entangled ground states}

The pair creation Hamiltonian is also found in spinor Bose-Einstein condensates (BECs) under the single-mode approximation, where each spin component has the same spatial wavefunction~\cite{Pu00}. In conjunction with a quadratic Zeeman shift parameterized by $q$, the Hamiltonian is given by
\begin{equation}
\hat{H} = q\hat{n}_0 + \frac{\Lambda}{N} \left( \hat{S}_x^2 + \hat{S}_y^2 \right).\label{eq:BECham}
\end{equation}
In the limit that $|q| \gg |\Lambda|$, the ground state of the Hamiltonian approaches the ground state of the quadratic Zeeman shift. If $q < 0$, this corresponds to all atoms in the $\ket{m=0}$ state, which can be prepared via optical pumping. In the limit that $|\Lambda| \gg q$, then the ground states become states with minimal (positive $\Lambda$) or maximal ($\Lambda < 0$) spin length meaning that both limits have large degenerate ground state manifolds. Of particular interest are the solutions with no magnetization: the macroscopic spin singlet $\ket{S=0,m=0}$ for positive $\Lambda$, and the symmetric Dicke state $\ket{S=N,m=0}$ for negative $\Lambda$. Both of these states are maximally entangled with Heisenberg limited metrological sensitivity. 

Adiabatic transformation of the Hamiltonian starting from the $\ket{m=0}^{\otimes N}$ prepares one of these highly entangled states~\cite{Zhang13}. Since the Hamiltonian conserves $\hat{S}_z$ it does not change the magnetization of the state. Unfortunately, the energy gap between the ground state and the first excited state at the transition point, $q=\pm4\Lambda$, closes as $N$ increases, scaling as $\Delta_{\mathrm{gap}}\sim N^{-1/3}$~\cite{Zhang13}. Note that this gap is calculated to the first excited state within the subspace that satisfies $\braket{\hat{S}_z}=0$, not arbitrary states. The gap is not strictly zero for finite numbers of atoms, but its size limits a true adiabatic sweep to a timescale far longer than the BEC lifetime~\cite{Hoang16PNAS} and sweeps in shorter timescales lead to poor overlap with the target state~\cite{Luo17} though still with enhanced metrological capabilities~\cite{Zou18}. Quantum control methods allow for improved fidelity without adiabaticity~\cite{Guo21,Xue22,Liu22,Mao23,Xin22,Xin23}, but do not allow for perfect fidelity even in theory.

\subsection{Rescaling the ground state energy gap}

A nematic term can be added to the spinor BEC Hamiltonian in Eq.~\eqref{eq:BECham} by interfacing the BEC with a cavity. Following the strategy described in Sec.~\ref{sec:nematicmixing} leads to an engineered Hamiltonian of the form
\begin{equation}
\hat{H} = q\hat{n}_0 + \frac{\Lambda}{N} \left( \hat{S}_x^2 + \hat{S}_y^2 \right)
+ \frac{\Phi}{N} \left( \hat{Q}_{xz}^2 + \hat{Q}_{yz}^2 \right). \label{eq:fullmixingham}
\end{equation}
Note that the spinor-BEC Hamiltonian can itself be emulated in a cavity~\cite{Masson17,Davis19}, though to engineer this entire Hamiltonian would require the use of two separate cavity modes as the frequency requirements are distinct for each of the two terms.

\begin{figure}
    \includegraphics[width=\columnwidth]{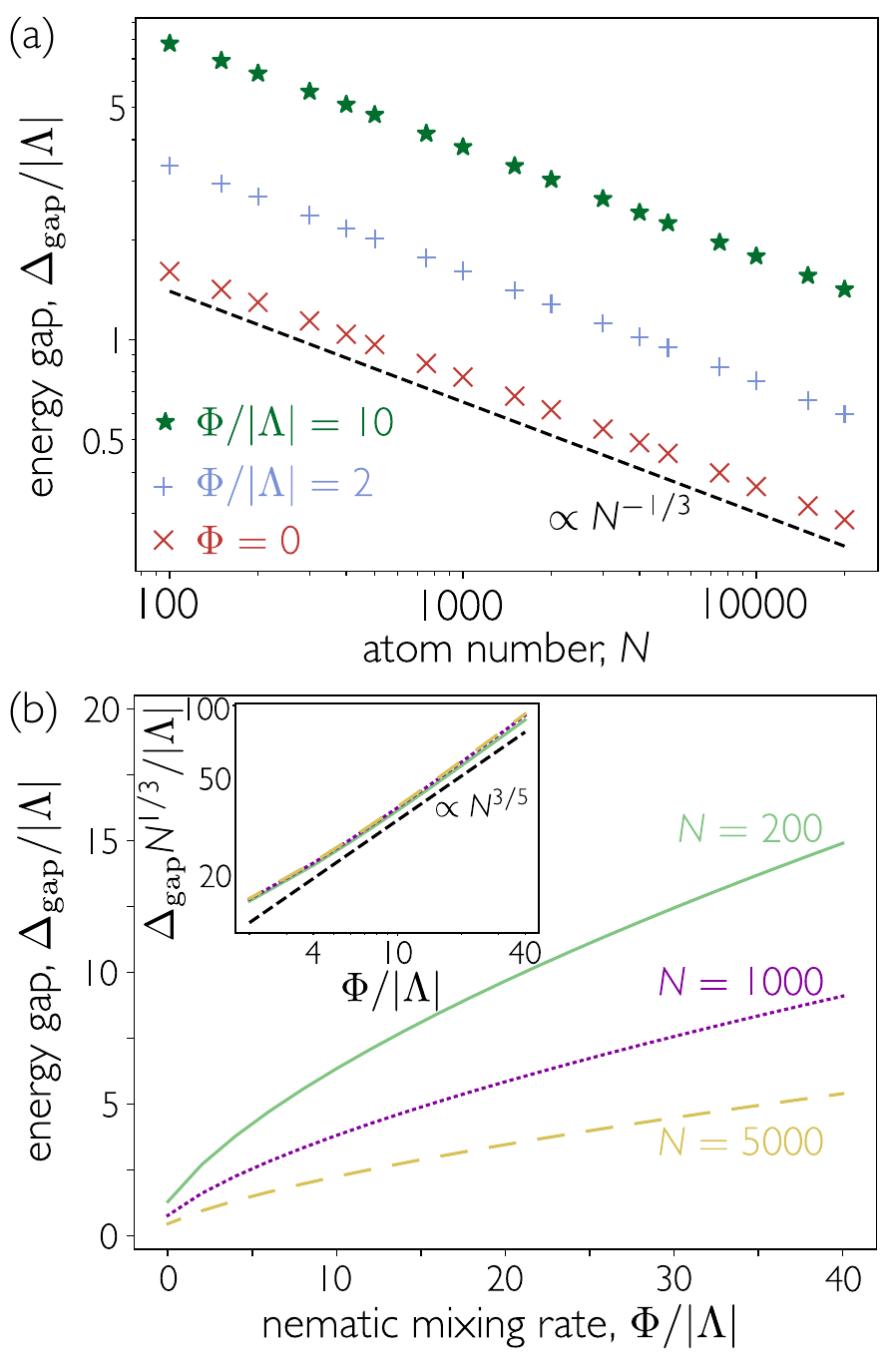}
    \caption{Spinor BEC energy gap with nematic mixing. (a)~Scaling of the energy gap with atom number with different nematic mixing rates. The dashed line is a guide-to-the-eye that scales as $N^{-1/3}$. (b) Scaling of the energy gap with nematic mixing rate for different atom numbers. The inset shows the same data rescaled by $N^{-1/3}$ with a guide-to-the-eye that scales as $N^{3/5}$. In both plots, $\Lambda$ is negative and the energy gap is found by minimizing the gap over quadratic Zeeman shift $q$ in the subspace that satisfies $\braket{\hat{S}_z} = 0$.\label{fig:gap}}
\end{figure}

\begin{figure*}
    \includegraphics[width=\textwidth]{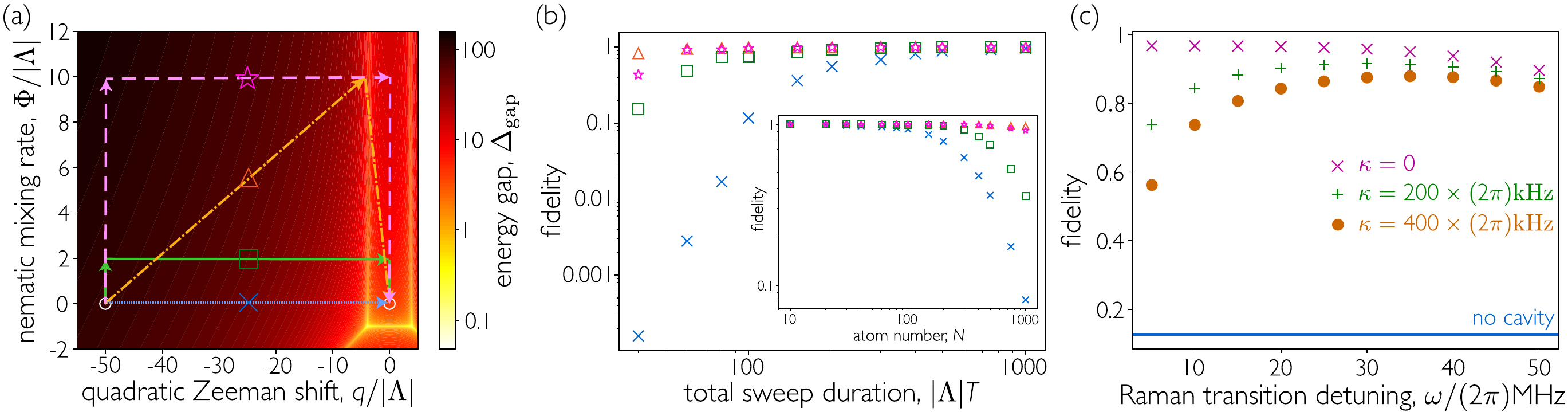}
    \caption{Fidelity of central Dicke state, $\ket{S=N,m=0}$, production via sweeping parameters. (a) Map of the minimum gap, $\Delta_{\mathrm{gap}}$, in parameter space for $N=500$ and $\Lambda$ taken to be negative. Arrowed paths show the parameter sweeps used in the other plots while the white circles show the start and end points. (b) Fidelity between the target state and the state produced by each of the four paths varying either the total duration of the sweep or (inset) atom number. The initial state is the instantaneous ground state of Eq.~\eqref{eq:fullmixingham} with $\set{q,\Phi}/|\Lambda| = \set{-50,0}$ and all sweeps are performed with linear variation in time. The paths are as follows: without nematic mixing turning off the quadratic Zeeman term (blue crosses); turning on the nematic mixing term over a time $0.01T$, turning off the quadratic Zeeman term over a time $0.9T$, and finally turning off the nematic mixing term over a time $0.09T$ for small (green squares) and large (pink stars) maximum nematic mixing rates; simultaneously ramping on the nematic mixing term while the quadratic Zeeman term is decreased to the phase transition point at $q=-4|\Lambda|$ over a time $0.5T$ and then simultaneously ramping off the nematic mixing term while the quadratic Zeeman term is further ramped to zero over $0.5T$. In (b) $N=500$ and in the inset $|\Lambda|T=100$. (c) Fidelity between the target state and the state produced via nematic sweeps with realistic cavity parameters. The ramp uses the orange path from (a) with  each half of the ramp lasting $|\Lambda| T= 2.5$ and $N=40$. Points with loss are obtained by solution to the master equation. The horizontal line is fidelity achieved by a sweep of the quadratic Zeeman shift without nematic mixing. \label{fig:gapscaling}}
\end{figure*}

The energy gap between the ground state and the first excited state is increased by the presence of the nematic term, as shown in Fig.~\ref{fig:gap}(a). The scaling of the energy gap still follows a scaling $\sim N^{-1/3}$, but a large nematic term can dramatically increase the prefactor. This means that adiabatic passage through the critical point can be much faster in finite-sized systems.

\subsection{State preparation}

Increasing the size of the gap allows for high-fidelity state preparation with a fully adiabatic parameter sweep. The parameter landscape of the energy gap is shown in Fig.~\ref{fig:gapscaling}(a). The standard sweep without using a nematic term passes through the phase transition at $q = -4|\Lambda|$ where the gap is smallest and must be extremely slow to achieve adiabaticity. An adiabatic route can instead be taken in the same time by using a small amount of the ramp time to turn on and off the nematic term. This ramp passes through the same phase transition, which remains at the same approximate value of $q=-4|\Lambda|$, in a parameter regime with a much larger gap. For the three ramps considered in Fig.~\ref{fig:gapscaling}(b), a fidelity >90\% for the central Dicke state, $\ket{S=N,m=0}$, is achieved for substantially shorter ramp times than is required without the cavity. Alternatively, one can view this as allowing for high fidelity production for substantially larger systems, as shown in the inset to Fig.~\ref{fig:gapscaling}(b).

Maximizing fidelity requires the nematic mixing term to be larger than the BEC terms. For an implementation interfacing the spinor BEC with a cavity directly, a typical order of magnitude for the collisional term is $1 \times 2\pi$ Hz for 40,000 atoms~\cite{Hoang16PNAS}. Taking a typical cavity coupling of $g = 1 \times 2\pi$ MHz, the cavity-drive detuning $\omega = 20 \times 2\pi$ MHz, and assuming $\kappa \ll \omega$, an estimate of the Rabi frequency required to produce a desired ratio of $\Phi$ to this collisional rate follows $\Omega \approx 2 \sqrt{\Phi/|\Lambda|} \times 2\pi \mathrm{MHz}$. This means that relatively little power is required to significantly increase the energy gap. Note that $\Phi$ scales as both $g^2$ and $\Omega^2$ so any increase to $g$ lowers the requirement on $\Omega$ by the same factor.

Realistic cavity parameters still allow for substantial improvement. As discussed in Sec.~\ref{sec:realistic}, it is not possible to achieve the nematic mixing Hamiltonian with both the cavity and laser at the required detuning. Realistic treatment of the cavity produces two potential sources of error for the adiabatic sweeps. Firstly, loss of photons through the cavity changes the magnetization and thus leads to states with zero overlap with the target state. Secondly, finite $\omega$ leads to terms that do not conserve $\hat{S}_z$, potentially opening up pathways to states with significantly smaller gaps. However, this seems to be an extremely minor effect in comparison to loss, as shown in Fig.~\ref{fig:gapscaling}(c). As such, relatively large values of $\omega$ offer extremely high fidelity and substantial improvement over a sweep without a cavity. It should be noted that, as in Sec.~\ref{sec:postselection}, the loss in fidelity due to cavity decay can also be mitigated via postselection.

Additional speedup and fidelity could be obtained by optimizing the ramp. The simple linear ramps used in Fig.~\ref{fig:gapscaling} are suboptimal because parameters change at the same rate regardless of the gap value. A more intelligent approach would have the ramp change quickly when the gap is large and slowly when the gap is small. These optimizations allow for the standard sweep, not using the nematic term, to be much faster than the linear ramp. The same speed-up would also be an option for the protocol described here. There is still a small loss in fidelity due to nonadiabatic effects in the nematic version, which could be minimized using quantum control methods in the same manner as for the standard sweep~\cite{Guo21, Liu22, Xin22, Mao23}. Together, these methods offer the possibility for rapid production of extremely high fidelity entangled states in spinor BECs interfaced with optical cavities.

\section{Outlook}

I have shown how to use multi-photon transitions in cavity QED to engineer interactions involving operators associated with the nematic tensor. Choosing a specific value of the detuning of driving lasers allows for an effective model where \textit{only} these nematic terms are non-zero. In the limit that the multi-photon transitions themselves are strongly off-resonant, the cavity mode can be adiabatically eliminated. This leads to a family of models of atom-photon interactions akin to Lipkin-Meshkov-Glick models~\cite{Lipkin65} but with nematic tensor operators rather than angular momentum operators. The nematic terms can be used for the generation of spin-nematic squeezing in the same manner as angular momentum terms, but at a significantly faster rate because of the accessibility of the detuning requirement. I estimate potential speed-ups for spin-nematic squeezing experiments of over a hundred times. The same terms also allow for rescaling of the energy gap of the spinor BEC collisional Hamiltonian, allowing for adiabatic preparation of highly entangled ground states.

The use of entire hyperfine manifolds in cavity QED has led to breakthroughs in interaction engineering and entanglement generation, but has so far been limited to exploiting the expanded state space rather than the associated operator space. The results of this work show that engineering atom-photon interactions in the expanded operator space can enhance cavity QED systems. On the one hand, results from the angular momentum case can be replicated but significantly more efficiently. On the other hand, these new terms can be used to provide their own unique physics. While the results presented here focussed on the particular case of the $F=1$ manifold of $^{87}$Rb driven on the $D_1$ line, the presence of the necessary condition should be ubiquitous for any alkali atom with an $F=1$ hyperfine manifold driven via either the $D_1$ or $D_2$ lines. Furthermore, Hamiltonians of the same form could be engineered for atoms in higher $F$ manifolds. Despite the extra states, engineered interactions are still described by the same operators as for spin-1 atoms but dynamics occur in a richer state space, potentially providing the opportunity for unique physics.

Harnessing the alternative forms of interactions available to spin-1 atoms offers significant potential for quantum simulation. While the focus of this work has been entanglement generation, it should be noted that these nematic terms could also be used to add additional complexity to Dicke models. The simple addition of a quadratic Zeeman term to the Dicke model Hamiltonian shows a rich phase diagram with regions of multistability, limit cycle solutions, and chaos~\cite{Lin22,Fan23,Adiv24,Adiv26}, motivating the study of how other nematic tensor operators would modify steady-state superradiant behavior. If the cavity interacts with the nematic degrees of freedom of the atoms, then it seems likely that superradiant phases associated to the nematic order should emerge, raising questions of the structure of the cavity field in such a state. If both nematic and spin terms are on equal footing then this should elicit competition between differently ordered phases. Combination of such systems with the ability to carefully control each atom's coupling to the cavity~\cite{Deist22,Liu23,Zhenjie23} may provide a rich phase diagram featuring high-order symmetry breaking~\cite{Ho26}.

\begin{acknowledgements}
I thank Ricardo Guti\'errez-J\'auregui for helpful comments and careful reading of the manuscript. This research was supported in part by grant NSF PHY-2309135 to the Kavli Institute for Theoretical Physics (KITP).
\end{acknowledgements}

\appendix

\section{Derivation of the atom-cavity effective master equation\label{app:excitedstates}}

I consider an ensemble of $N$ $^{87}$Rb atoms coupled to a high-finesse optical cavity driven off-resonantly on the $D_1$ line. The master equation governing evolution of this system is
\begin{widetext}
\begin{equation}
\dot{\rho} = - \ii \left[\sum\limits_{j=1}^N \hat{H}_0^j + \hat{H}_\mathrm{int.}^j,\rho\right] + \sum\limits_k \kappa_k\left(2 \hat{a}_k \rho \hat{a}_k^\dagger - \hat{a}_k^\dagger\hat{a}_k \rho - \rho \hat{a}_k^\dagger\hat{a}_k\right)
\end{equation}
\end{widetext}
where I consider a set of cavity modes with annihilation operator $\hat{a}_k$ and decay rate $\kappa_k$. It is assumed that atomic decay can be ignored due to negligible excited state population. The Hamiltonian sums over all atoms internal energy and their interaction with the field, but there is no direct interaction between atoms. This means we can perform the adiabatic elimination for each atom independently and then sum over these Hamiltonians. As such, we now drop the $j$ notation from each Hamiltonian and consider only a single atom. 

The first term of the Hamiltonian governs the energy of the atomic states and is given by
\begin{widetext}
\begin{equation}
    \hat{H}_0^j = \sum\limits_{m = -1}^{1} m \omega_{z} \ket{m}_j \bra{m}_j + \sum\limits_{F=1,2}\sum\limits_{f = -F}^{F} \omega_{F} \ket{F,f}_j\bra{F,f}_j.
\end{equation}
\end{widetext}
Here, I adopt notation that the $F=1$ ground magnetic states of the $j$th atom are labeled $\ket{m}_j$, while excited states of the same atom are labeled with both the hyperfine level and magnetic number $\ket{F,f}_j$. The ground states have frequencies governed by a linear Zeeman shift $m \omega_z$, while $\omega_F$ is the frequency of the $F$ excited state hyperfine manifold. Differences between the energies of the magnetic states within each excited state manifold are assumed to be negligible in comparison to the detuning and can thus be ignored.

The interaction Hamiltonian is given by
\begin{widetext}
\begin{equation}
 \hat{H}_{\mathrm{int.}} = \sum\limits_{m,F,f} \sum\limits_{\mu = \sigma_-,\pi,\sigma_+} \hat{\Omega}_\mu(t) d^{F}_{m,q_\mu} \ket{F,f+q_\mu}\bra{m} + \hat{\Omega}^\dagger_\mu(t) d^{F}_{m,q_\mu} \ket{m}\bra{F,f+q_\mu}
 \end{equation}
 \end{widetext}
 where the first sum runs over all states, and the second over polarization components $\mu$. Each polarization has associated changes in magnetization $q_\mu = \set{-1,0,+1}$ for $\mu = \set{\sigma_-,\pi,\sigma_+}$ and $d^{F}_{m,q_\mu}$ is the dipole matrix element for the transition $\ket{m} \leftrightarrow \ket{F,f+q_\mu}$ expressed in units of $\bra{J=1/2}|e\hat{r}|\ket{J'=1/2}$ where $e\hat{r}$ is the dipole operator. I also introduce ``driving'' operators for each polarization component which correspond to summations over classical drives and cavity modes with that polarization. For example, if for a particular polarization $\mu$ there is one laser drive at frequency $\omega_l$ and one cavity mode with frequency $\omega_c$ then this operator is
 \begin{equation}
     \hat{\Omega}_\mu(t) = \frac{\Omega}{2} \mathrm{e}^{-\ii \omega_l t} \mathbb{1} + g \mathrm{e}^{-\ii \omega_c t} \hat{a}
 \end{equation}
 where $\Omega$ is the Rabi frequency of the driving laser and $g$ the single-atom cavity coupling rate. While this operator is time-dependent, for brevity the dependence on $t$ will be omitted henceforth.

 \subsection{Adiabatic elimination of the excited states}

 Adiabatic elimination of the excited states can be performed using a projective operator formalism~\cite{Reiter12} leading to an effective interaction Hamiltonian with generic form
 \begin{widetext}
     \begin{align}
         \hat{H}_{\mathrm{eff.}} &= \left(\sum\limits_{m,F,f,\mu} \hat{\Omega}^\dagger_\mu  d^{F}_{m,q_\mu} \ket{m} \bra{F,f+q_\mu} \right) \left( \sum\limits_{G,g}\frac{ \ket{G,g}\bra{G,g}}{\Delta_G} \right) \left(\sum\limits_{n,H,h,\nu} \hat{\Omega}_\nu  d^{H}_{h,q_\nu}\ket{H,h+q_\nu} \bra{n}  \right) \\
         &= \sum\limits_{\mu,\nu} \sum\limits_{n,m} \sum\limits_{F} \delta_{m+q_\nu,n+q_\mu} \frac{\hat{\Omega}^\dagger_\mu \hat{\Omega}_\nu  d^{F}_{m,q_\mu} d^{F}_{n,q_\nu}}{\Delta_F} \ket{m} \bra{n}\label{eq:adiabaticelimsum}
     \end{align}
 \end{widetext}
 where $\delta_{m+q_\nu,n+q_\mu}$ is a Kronecker delta that ensures there is a two-photon pathway to connect the states $\ket{m}$ and $\ket{n}$ and $\Delta_F = \omega_{c,l} - \omega_F$ is the detuning of the fields from the hyperfine level $F$. Note that here I have assumed that all laser and cavity frequencies are equal. This assumption is for the sake of brevity and is not necessary to the elimination. To undo this approximation, one makes the following substitution to all terms~\cite{Reiter12}
 \begin{equation}
     \frac{1}{\Delta_F} \rightarrow \frac{(\omega_1 + \omega_2 - 2\omega_F)}{2(\omega_1 - \omega_F)(\omega_2 - \omega_F)}
 \end{equation}
 where $\omega_1$ and $\omega_2$ are the frequencies of the two contributing frequency terms.

Each term in the sum in Eq.~\ref{eq:adiabaticelimsum} represents a two-photon process where a photon of polarization $\mu$ is absorbed by the atom and a photon of polarization $\nu$ is emitted. There are two categories of process that we will consider. Firstly, if the photons are of the same polarization then $m=n$ and the process does not change the state. These processes will lead to ac Stark shifts and dispersive shifts of the atomic frequencies. Secondly, if the photons are not of the same polarization then $m\neq n$ and the term describes a Raman transition. Raman transitions can further be split into categories where one photon is $\pi$-polarized and the magnetic state changes by one or neither photon is $\pi$-polarized and the magnetic state changes by two.

The Hamiltonian can be split into terms proportional to $\ket{m}\bra{n}$, which I denote as $\hat{H}_{m,n}$, dropping the effective label for brevity. The set of these terms corresponding to two photons of the same polarization are
\begin{widetext}\begin{subequations}
\begin{align}
    \hat{H}_{-1,-1} &= \left[ \frac{\hat{\Omega}^\dagger_{-} \hat{\Omega}_- }{2\Delta_2} + \frac{\hat{\Omega}^\dagger_{\pi} \hat{\Omega}_\pi }{12\Delta_1} + \frac{\hat{\Omega}^\dagger_{\pi} \hat{\Omega}_\pi }{4\Delta_2} + \frac{\hat{\Omega}^\dagger_{+} \hat{\Omega}_+ }{12\Delta_1} + \frac{\hat{\Omega}^\dagger_{+} \hat{\Omega}_+ }{12\Delta_2} \right] \ket{-1}\bra{-1} \\
    \hat{H}_{0,0} &= \left[ \frac{\hat{\Omega}^\dagger_{-} \hat{\Omega}_- }{12\Delta_1} + \frac{\hat{\Omega}^\dagger_{-} \hat{\Omega}_- }{4\Delta_2} + \frac{\hat{\Omega}^\dagger_{\pi} \hat{\Omega}_\pi }{3\Delta_2} + \frac{\hat{\Omega}^\dagger_{+} \hat{\Omega}_+ }{12\Delta_1} + \frac{\hat{\Omega}^\dagger_{+} \hat{\Omega}_+ }{4\Delta_2} \right] \ket{0}\bra{0} \\
    \hat{H}_{1,1} &= \left[ \frac{\hat{\Omega}^\dagger_{-} \hat{\Omega}_- }{12\Delta_1} + \frac{\hat{\Omega}^\dagger_{-} \hat{\Omega}_- }{12\Delta_2} + \frac{\hat{\Omega}^\dagger_{\pi} \hat{\Omega}_\pi }{12\Delta_1} + \frac{\hat{\Omega}^\dagger_{\pi} \hat{\Omega}_\pi }{4\Delta_2} + \frac{\hat{\Omega}^\dagger_{+} \hat{\Omega}_+ }{2\Delta_2} \right] \ket{1}\bra{1}.
\end{align}
    \end{subequations}
These three terms can be written as a combination of the spin magnetization operator $\hat{\sigma}_z = \ket{1}\bra{1} - \ket{-1}\bra{-1}$, its square $\hat{\sigma}_{z}^2 = \ket{1}\bra{1} + \ket{-1}\bra{-1}$, and the identity $\mathbb{1} = \ket{1}\bra{1} + \ket{0}\bra{0} + \ket{-1}\bra{-1}$. This results in
\begin{align}
  \hat{H}_{-1,-1} + \hat{H}_{0,0} + \hat{H}_{1,1} = \left(\frac{1}{24\Delta_1} - \frac{5}{24\Delta_2}\right)&\left(\hat{\Omega}^\dagger_{-} \hat{\Omega}_-  - \hat{\Omega}^\dagger_{+} \hat{\Omega}_+ \right) \hat{\sigma}_z+  \left(\frac{1}{12\Delta_1} + \frac{1}{4\Delta_2} \right) \left(\hat{\Omega}^\dagger_{-} \hat{\Omega}_-  + \hat{\Omega}^\dagger_{+} \hat{\Omega}_+ \right) \\&+ \frac{\hat{\Omega}^\dagger_{\pi} \hat{\Omega}_\pi }{3\Delta_2}
  + \left(\frac{1}{24\Delta_2} - \frac{1}{24\Delta_1} \right)\left( \hat{\Omega}^\dagger_{-} \hat{\Omega}_-  + \hat{\Omega}^\dagger_{+} \hat{\Omega}_+  - 2\hat{\Omega}^\dagger_{\pi} \hat{\Omega}_\pi \right)\hat{\sigma}_z^2
\end{align}

The next pair of terms corresponds to increasing the magnetic state value by one. This can be done either by absorbing a $\pi$ polarized photon and emission of a $\sigma_-$ polarized photon, or by absorbing a $\sigma_+$ polarized photon and emission of a $\pi$ polarized photon. The Hamiltonian terms are given by
\begin{subequations}
\begin{align}
    \hat{H}_{0,-1} &= \left[ -\frac{\hat{\Omega}^\dagger_\pi\hat{\Omega}_+}{6\Delta_2} + \frac{\hat{\Omega}^\dagger_-\hat{\Omega}_\pi}{12\Delta_1} - \frac{\hat{\Omega}^\dagger_-\hat{\Omega}_\pi}{4\Delta_2}\right] \ket{0} \bra{-1} \\
    \hat{H}_{1,0} &= \left[\frac{\hat{\Omega}^\dagger_\pi\hat{\Omega}_+}{12\Delta_1} - \frac{\hat{\Omega}^\dagger_\pi\hat{\Omega}_+}{4\Delta_2} - \frac{\hat{\Omega}^\dagger_-\hat{\Omega}_\pi}{6\Delta_2}\right] \ket{1} \bra{0}
\end{align}
\end{subequations}
which can be written as a combination of the spin raising operator $\hat{\sigma}_+ = \sqrt{2} \left( \ket{0}\bra{-1} + \ket{1}\bra{0}\right)$ and a nematic version which resembles a twisted raising operator due to a sign difference on the two terms $\hat{q}_{+z} = \hat{q}_{xz} + \ii \hat{q}_{yz} = 2\sqrt{2} \left( -\ket{0}\bra{-1} + \ket{1}\bra{0} \right)$. Using this we can write
\begin{equation}
\hat{H}_{0,-1} + \hat{H}_{1,0} = \left(\frac{1}{24\sqrt{2}\Delta_1} - \frac{5}{24\sqrt{2}\Delta_2} \right) \left( \hat{\Omega}_\pi^\dagger \hat{\Omega}_+ + \hat{\Omega}_-^\dagger\hat{\Omega}_\pi \right)\hat{\sigma}_+ + \left(\frac{1}{48\sqrt{2}\Delta_1} - \frac{1}{48\sqrt{2}\Delta_2} \right) \left( \hat{\Omega}_\pi^\dagger\hat{\Omega}_+ - \hat{\Omega}_-^\dagger\hat{\Omega}_\pi \right) \hat{q}_{+z}.
\end{equation}
\end{widetext}
The effective Hamiltonian must be Hermitian, providing the forms of two other Raman transition terms $\hat{H}_{-1,0} = \hat{H}_{0,-1}^\dagger$ and $\hat{H}_{0,1} = \hat{H}_{1,0}^\dagger$.

The last pair of terms correspond to changing the magnetic state by two and are calculated as
\begin{subequations}
    \begin{align}
        \hat{H}_{1,-1} =  \left( \frac{1}{12\Delta_1} - \frac{1}{12\Delta_2}\right) \hat{\Omega}^\dagger_- \hat{\Omega}_+ \ket{1}\bra{-1} \\
        \hat{H}_{-1,1} = \left( \frac{1}{12\Delta_1} - \frac{1}{12\Delta_2}\right) \hat{\Omega}^\dagger_+ \hat{\Omega}_- \ket{-1}\bra{1}.
    \end{align}
\end{subequations}
The coherence operators here can be written as combinations of $\hat{q}_{xx}$, $\hat{q}_{yy}$ and $\hat{q}_{xy}$ but a simpler notation arises using that $\hat{\sigma}_+^2 = 2 \ket{1}\bra{-1}$ and $\hat{\sigma}_-^2 = 2 \ket{-1}\bra{1}$.

To find the particular form of this effective interaction Hamiltonian, one substitutes in the forms of the field operators, sums over $\hat{H}_{m,n}$ and sums over all $N$ atoms.

\subsection{Derivation of the scheme used for entanglement generation\label{app:entanglementscheme}}

The entanglement described in the main text is generated using a single laser drive that is circularly-polarized and a single $\pi$-polarized cavity mode. This means that, choosing arbitrarily for a $\sigma_-$-polarized drive as the circularly polarization, that the fields are
    \begin{align}
        \hat{\Omega}_\pi = g \mathrm{e}^{-\ii\omega_c t} \hat{a}; ~~~~ \hat{\Omega}_- = \frac{\Omega}{2} \mathrm{e}^{-\ii\omega_l t} \mathbb{1}; ~~~~
        \hat{\Omega}_+ = 0
    \end{align}
Substituting in this choice of fields leads to
\begin{widetext}
    \begin{align}
        \hat{H}_{\mathrm{eff.}} = &\left[\omega_z + \Omega^2 \left(\frac{1}{96\Delta_1} - \frac{5}{96\Delta_2}\right) \right] \hat{\sigma}_z + g^2\left(\frac{1}{18\Delta_1} + \frac{5}{18\Delta_2} \right) \hat{a}^\dagger\hat{a}
  + \left( \frac{\Omega^2}{4}  - 2g^2\hat{a}^\dagger\hat{a}\right)\left(\frac{1}{48\Delta_2} - \frac{1}{48\Delta_1} \right)\hat{q}_{zz} \notag \\&+ \frac{\Omega g}{2}\left(\frac{1}{24\sqrt{2}\Delta_1} - \frac{5}{24\sqrt{2}\Delta_2} \right)   \left(\hat{a} \mathrm{e}^{-\ii\left(\omega_c - \omega_l\right)t} \hat{\sigma}_+ + \hat{a}^\dagger \mathrm{e}^{-\ii\left(\omega_l-\omega_c\right)t}\hat{\sigma}_- \right)\notag \\&+ \frac{\Omega g}{2} \left(\frac{1}{48\sqrt{2}\Delta_2} - \frac{1}{48\sqrt{2}\Delta_1} \right) \left(\hat{a}\mathrm{e}^{-\ii\left(\omega_c-\omega_l\right)t} \hat{q}_{+z} + \hat{a}^\dagger\mathrm{e}^{-\ii\left(\omega_l-\omega_c\right)t} \hat{q}_{-z} \right)
    \end{align}
\end{widetext}
where $\omega_z$ is the linear Zeeman shift, I have used that $\hat{\sigma}_{z}^2 = \hat{q}_{zz}/2 + 2/3 \mathbb{1}$, and dropped all terms that are only proportional to the identity.

For $N$ atoms, this Hamiltonian must be summed over each atom. Due to the permutational symmetry, the result is simply to transform the single atom operators into collective operators. This leads to the form provided in the main text
\begin{widetext}
    \begin{align}
        \hat{H}_{\mathrm{eff.}} = \omega_0 \hat{S}_z + \omega \hat{a}^\dagger\hat{a} + \omega_q \hat{Q}_{zz} + \frac{\delta_q}{2N} \hat{Q}_{zz} \hat{a}^\dagger\hat{a} + \frac{\lambda}{\sqrt{2N}} \left( \hat{a}\hat{S}_+ + \hat{a}^\dagger\hat{S}_- \right) + \frac{\phi}{2\sqrt{2N}} \left( \hat{a}\hat{Q}_{+z} + \hat{a}^\dagger\hat{Q}_{-z} \right)
    \end{align}
\end{widetext}
with effective parameters
\begin{subequations}
    \begin{align}
        \omega_0 &= \omega_z + \frac{\Omega^2}{96} \left(\frac{1}{\Delta_1} - \frac{5}{\Delta_2}\right) \\
        \omega &=  \omega_c - \omega_l + \frac{N g^2}{18}\left(\frac{1}{\Delta_1} + \frac{5}{\Delta_2} \right)  \\
        \omega_q &=  \frac{\Omega^2}{192}\left(\frac{1}{\Delta_2} - \frac{1}{\Delta_1} \right) \\
        \delta_q &= \frac{Ng^2}{12} \left( \frac{1}{\Delta_1} - \frac{1}{\Delta_2} \right) \\
        \lambda &= \frac{\sqrt{N}\Omega g}{48}\left(\frac{1}{\Delta_1} - \frac{5}{\Delta_2} \right)  \\
        \phi &= \frac{\sqrt{N} \Omega g}{48} \left(\frac{1}{\Delta_2} - \frac{1}{\Delta_1} \right)
    \end{align}
\end{subequations}
where the time-dependence has been removed by moving into a rotating cavity frame.

\bibliography{bibliography.bib}

\end{document}